\documentclass[12pt]{article}
\usepackage{amssymb,amsmath,amsthm,amsfonts,amscd}
\newcommand\be{\begin{equation}}
\newcommand\ee{\end{equation}}
\newcommand\bea{\begin{eqnarray}}
\newcommand\eea{\end{eqnarray}}

\newcommand{\fatalpha}{{\bf \alpha \kern -0.44em \alpha}}
\newcommand{\fatsigma}{{\bf \sigma \kern -0.54em \sigma}}
\newcommand{\tpchi}{{\bf \chi \kern -0.35em \chi}}
\newcommand{\llambda}{{\bf \lambda \kern -0.45em \lambda}}

\bibliography{plain}
\title{\bf Generalized Grassmannian  Coherent States For Pseudo-Hermitian $n$ Level Systems } 
\author{ G. Najarbashi $^{a}$
 \thanks{E-mail:najarbashi@uma.ac.ir} ,
M. A. Fasihi $^{b,c}$
\thanks{E-mails:fasihi@alice.math.kindai.ac.jp}, H. Fakhri $^{d}$
 \thanks{E-mail:hfakhri@tabrizu.ac.ir } \\
\\ $^a${\small Department of Physics, University of Mohaghegh Ardabili,
Ardabil 179, Iran.}\\$^b${\small Research Center for Quantum Computing, Interdisciplinary Graduate School of Science}\\{\small and Engineering, Kinki University, Higashi-Osaka, Osaka 577-8502, Japan.}\\
$^c${\small Department of Physics, Azarbijan University of Tarbiat
Moallem, Tabriz 53714-161,Iran.}\\
$^d${\small Department of Theoretical Physics and
Astrophysics, Tabriz University, Tabriz 51664, Iran.}} \pagebreak

\begin{document}
\maketitle
\newpage 

\begin{abstract}
The purpose of this paper is to generalize fermionic  coherent
states for two-level systems described by pseudo-Hermitian
Hamiltonian \cite{Trifonov}, to n-level systems. Central to this
task is the expression of the coherent states in terms of
generalized Grassmann variables. These kind of Grassmann coherent
states  satisfy bi-overcompleteness condition instead of
over-completeness one, as it is reasonably expected because of the
biorthonormality of the system. Choosing an appropriate Grassmann
weight function resolution of identity is examined. Moreover
Grassmannian coherent and squeezed states of deformed group
$SU_{q}(2)$  for three level pseudo-Hermitian system are presented.
\\
{\bf PACs Index: 03.65.-w, 03.65.Ca}
\end{abstract}
\pagebreak

\vspace{7cm}
\section{Introduction}
 The last decade have witnessed a growing interest in
non-Hermitian Hamiltonians with real spectra,
\cite{Scholtz,Ben1,Ben2,Ben3,Ben4,Can,Znojil}. Considering  the results of
various numerical studies, Bender and his collaborators
\cite{Ben2,Ben3} found certain examples of one-dimensional
non-Hermitian Hamiltonians that possessed real spectra. Because
these Hamiltonians were invariant under $PT$ transformations, their
spectral properties were linked with their $PT$-symmetry. Later
Mostafazadeh introduced the notion of pseudo-Hermiticity as an
alternative possible approach for a non-Hermitian operator to admit
a real spectrum \cite{Mos1,Mos2}.
\par
 On the other hand Grassmannian
coherent states and their generalization
\cite{Kerner,Filip,Majid,Isa,Cug,ILInski} has attracted a great deal
of attention in the last decade and the concept of coherent states
was also introduced to $PT$ symmetric quantum mechanics, \cite{Bag,
Roy}, and pseudo-Hermitian one, \cite{Trifonov}. Here our objective
is to construct generalized Grassmannian pseudo-Hermitian coherent
states (GPHCS) by introducing lader operators for a general  $n$
level pseudo-Hermitian system and applying bi-orthonormal property
of this system to facilitate the investigation of over-completeness
of GPHCS. We find throughout this work that  many close parallels
can be established between the expressions evaluated for GPHCS and
the more familiar ones for boson coherent states. For boson coherent
states the integration for resolution of identity is taken over
commuting complex variables, while for fermions, on the other hand,
the integration is over anticommuting Grassmann numbers that have no
classical analogues. This is due to the fact that, in the context of
quantum field theory \cite{Ryder}, fermion fields  anti-commute,
hence their eigenvalues must, as noted by Schwinger be
anti-commuting numbers \cite{Schwinger}.
The  GPHCS may be useful  to describe entangled coherent states for compound systems
governed by non-Hermitian Hamiltonians in quantum information theory \cite{Najar1,Najar2}.
It is also possible to generalize P-function,
Q-function and Wigner function for fermions \cite{Glauber}, to
the pseudo-Hermitian density matrix of $n$ level systems \cite{Scolarici}.
\par
The paper is organized as follows. In Section 2, we introduce the
concept of a pseudo-Hermitian operator and consider the basic
spectral properties of pseudo-Hermitian Hamiltonians that have a
complete biorthonormal eigenbasis. In Section 3, we present the
generalized Grassmannian variables based on Majid approach
\cite{Majid} and define pseudo-Hermitian lader operators to
construct GPHCS. We show that unlike the canonical Hermitian
coherent state, GPHCS satisfy bi-over-completeness condition instead
of over-completeness one. We also study the stability of the GPHCS
and finally following the introduced approach we construct the GPHCS
for $SU_{q}(2)$.
\section{Pseudo-Hermitian Hamiltonians and Biorthonormal Eigenbasis}
Intensive study of Schrodinger equation with complex potentials, but with real spectrum, was performed by different methods. The pioneer papers \cite{Ben1, Ben2}, initiated investigation of $\mathrm{PT}$ symmetric systems and afterwards more general class of pseudo-Hermitian models was introduced by Mostafazadeh \cite{Mos1}.
Following the second approach, let $H : \mathcal{H}
\rightarrow \mathcal{H}$ be a linear operator acting in a Hilbert
space $\mathcal{H}$ and $\eta : \mathcal{H} \rightarrowtail
\mathcal{H}$ be a linear Hermitian automorphism (invertible
transformation). Then the $\eta$-pseudo-Hermitian adjoint of H is
defined by
\begin {equation}\label{ps1}
H^{\sharp} = \eta^{-1}H^\dag\eta.
\end{equation}
$H$ is said to be pseudo-Hermitian with respect to $\eta$ or
simply $\eta$-pseudo-Hermitian if $H^{\sharp} = H$. As in Refs.
\cite{Mos1,Mos2} the eigenvalues of pseudo-Hermitian Hamiltonian
$H$ are either real or come in complex-conjugate pairs and the
following relations in nondegenerate case  hold:
\begin {equation}\label{pseudo}
    H^{\dag}=\eta H \eta^{-1}.
\end{equation}
According to \cite{Mos1} we consider only diagonalizable operators
$H$ with discrete spectrum. Then, a complete biorthonormal
eigenbasis ${|\psi_{i}\rangle, |\phi_{i}\rangle}$ exist, i.e., a
basis such that
\begin {equation}\label{Hamiltonian}
    \begin{array}{c}
  H|\psi_{i}\rangle=E_{i}|\psi_{i}\rangle,\quad
H^{\dag}|\phi_{i}\rangle=\bar{E}_{i}|\phi_{i}\rangle, \\
  \langle\phi_{i}|\psi_{j}\rangle=\delta _{ij}, \\
  \sum_{i}|\psi_{i}\rangle\langle\phi_{i}|=\sum_{i}|\phi_{i}\rangle\langle\psi_{i}|=I. \\
\end{array}
\end{equation}
For a given pseudo-Hermitian $H$ there are infinitely many $\eta$
satisfying Eq.(\ref{pseudo}). These can however be expressed in
terms of a complete biorthonormal basis ${{|\psi_{i}\rangle,
|\phi_{i}\rangle}}$ of $H$. In non degenerate case the general
linear, Hermitian, invertible operator $\eta$ and it's inverse
satisfying Eq.(\ref{pseudo}) have the forms
\begin {equation}\label{eta}
    \eta=\sum_{i}|\phi_{i}\rangle\langle\phi_{i}|,\qquad \eta^{-1}=\sum_{i}|\psi_{i}\rangle\langle\psi_{i}|
\end{equation}
\begin {equation}\label{phipsietamenha}
    |\phi_{i}\rangle=\eta|\psi_{i}\rangle,\qquad
    |\psi_{i}\rangle=\eta^{-1}|\phi_{i}\rangle,
\end{equation}
where here we consider the non-Hermitian Hamiltonians with real spectrums, hence $\eta$ is positive definite operator.
\section{Generalized Grassmannian Pseudo-Hermitian Coherent States}
\subsection{Generalized Grassmannian variables}
The basic properties of generalized Grassmann variables are
discussed in Refs. \cite{Kerner,Filip,Majid,Isa,Cug,ILInski}. Here
we survey the properties that we shall make use of in article.
According to Ref \cite{Majid}, $Z_{n}$ graded  Grassmann algebra is
generated by the variables satisfying, by definition, the following
properties:
\begin {equation}\label{qdeform1}
    \begin{array}{c}
    \theta_{i}\theta_{j}=q\ \theta_{j}\theta_{i}\quad, \quad
i,j=1,2,...\;\; i<j \\
    \hspace{-3cm}\theta_{i}^n=0 , \qquad q=e^{\frac{2\pi i}{n}}.\\
  \end{array}
\end{equation}
Analogous rules also apply for the Hermitian conjugate of
$\theta$, $\theta^\dag=\bar{\theta}$, as:
\begin {equation}\label{qdeform2}
    \begin{array}{c}
    \bar{\theta}_{i}\bar{\theta}_{j}=q\ \bar{\theta}_{j}\bar{\theta}_{i}\quad, \quad
 i<j \\
   \hspace{-3cm} \bar{\theta}_{i}^n=0 .\\
  \end{array}
\end{equation}
Consider the generalized Berezin's rules of integration as :
\begin {equation}\label{brezin}
    \int d\theta \theta^{k}= \int d\bar{\theta}
\bar{\theta}^{k}=\delta_{k,n-1},
\end{equation}
where $k$ is any positive integer. We also need the following
relations which are necessary to compute the integral of any
function over the Grassmann algebra.
\begin {equation}\label{gG}
     \left\{\begin{array}{cc}
  \theta d \bar{\theta} = q\ d \bar{\theta} \theta \quad, & \bar{\theta} d \theta = q \ d \theta \bar{\theta} \\
  \theta d \theta = \bar{q}\ d \theta \theta \quad, & \bar{\theta} d \bar{\theta} = \bar{q}\ d \bar{\theta} \bar{\theta} \\
  d \theta d \bar{\theta} = \bar{q}\ d \bar{\theta} d\theta\quad, & \  \theta  \bar{\theta} = \bar{q}\  \bar{\theta} \theta \\
\end{array}\right.
\end{equation}
Now we are ready to give a prescription for the construction of
the generalized Grassmannian pseudo-Hermitian coherent state.
\subsection{Coherent States }
Now we will develop a formalism to construct the generalized
Grassmannian coherent states (GCS) for pseudo-Hermitian $n$ level
systems. Let introduce the generalized coherent states as the
eigen-states of annihilation operator $b$ which is defined as:
\begin {equation}\label{b}
b:=\sum_{i=0}^{n-1}\sqrt{\rho_{i+1}}\
|\psi_{i}\rangle\langle\phi_{i+1}|.
\end {equation}
Using Eq. (\ref{Hamiltonian}), it is straightforward to check that
the action of annihilation operator $b$ on $|\psi_{i}\rangle$
eigen-states is:
\begin {equation}\label{baction}
b|\psi_{i}\rangle=\sqrt{\rho_{i}}\ |\psi_{i-1}\rangle,
\end {equation}
so $|\psi_{0}\rangle$ is the vacuum state, and the annihilation
operator $b$ has the nilpotency degree of order $n$ i.e., $b^n=0$.
 Let have the following quantization relations
between the biorthonormal eigen-state \{${|\psi_{i}\rangle},\
{|\phi_{i}\rangle},\ i=0,1,2,...n-1$\} and generalized
Grassmannian variables $\theta, \bar{\theta}$
\begin{equation}\label{thetapsiphi}
\left\{\begin{array}{cc}
 \theta \ |\psi_{i}\rangle = q^{^{i-1}} \ |\psi_{i}\rangle\
 \theta \,\,\ & \ \bar{\theta}\ \langle\psi_{i}|\ = q^{^{i-1}} \ \langle\psi_{i}|\ \bar{\theta} \ \\
  \theta \ \langle\psi_{i}|\ = \bar{q}^{^{i-1}} \
\langle\psi_{i}|\ \theta \,\,\ & \ \bar{\theta} \ |\psi_{i}\rangle
\ = \bar{q}^{^{i-1}} \ |\psi_{i}\rangle\
\bar{\theta} \ \\
  \theta \ |\phi_{i}\rangle \ =q^{^{i-1}} \ |\phi_{i}\rangle \
 \theta\,\,\ &  \bar{\theta} \ \langle\phi_{i}| \ = q^{^{i-1}} \  \langle\phi_{i}|\ \bar{\theta}
 \\
  \theta \ \langle\phi_{i}|\ = \bar{q}^{^{i-1}} \
\langle\phi_{i}| \ \theta \,\,\ & \ \bar{\theta} \
|\phi_{i}\rangle \ = \bar{q}^{^{i-1}} \ |\phi_{i}\rangle \
\bar{\theta}.
\end{array}\right.
\end{equation}
Considering the Eqs. (\ref{eta}), (\ref{thetapsiphi}) it is easy to
check that
\begin {equation}\label{etateta}
[\theta,\eta]=[\theta,\eta^{-1}]=[\bar{\theta},\eta]=[\bar{\theta},\eta^{-1}]=0.
\end{equation}
The generalized Grassmannian pseudo-Hermitian coherent states (GPHCS) denoted by $|\theta\rangle_{n}$, by definition are the eigen-states of
annihilation operator $b$ with the eigen-values given by the label
of the coherent states, so we must find the GPHCS
$|\theta\rangle_{n}$ such that
\begin {equation}\label{coherent}
b\ |\theta\rangle_{n} =\theta \ |\theta\rangle_{n},
\end {equation}
where the eigenvalue $\theta$ is a complex generalized
Grassmannian variable. We write generically
\begin{equation}\label{coherentform}
|\theta\rangle_{n}=\sum_{i=0}^{n-1}\
 \alpha_{i}\theta^{i}\ |\psi_{i}\rangle,
\end{equation}
considering Eq.(\ref{coherent}) we get
\begin{equation}\label{coherentform1}
|\theta\rangle_{n}=\sum_{i=0}^{n-1}\
\frac{\bar{q}^{^{\frac{i(i+1)}{2}}}}{\sqrt{\rho_{i}!}}\ \theta
^{i}\ |\psi_{i}\rangle,
\end{equation}
where \{$\rho_{n}!:= \rho_{0}\rho_{1}\rho_{2}...\rho_{n}, \
\rho_{0}=1$\} and $\rho_{i}$s in general are complex variable.
\par
As it is usual for Hermitian systems, in order to express  the
states $|\theta\rangle_{n}$ in terms of the vacuum state we use
the Hermitian conjugate of the annihilation operator, $b^\dag$.
But it is important to not that, here we are dealing with
pseudo-Hermitian system, so as a result of this fact it is easy to
show that $b^\dag$ is not the creation operator for
$|\psi_{i}\rangle$ eigen-basis . To overcome this problem we need
to introduce the new operator $b^{\sharp}$, which is the $\eta$
pseudo-Hermitian of the $b^\dag$, as
\begin{equation}\label{bsharp1}
b^{\sharp}:= \eta^{-1}b^{\dag}\eta=\sum_{i=0}^{n-1}
\sqrt{\rho_{i+1}} |\psi_{i+1}\rangle\langle\phi_{i}|,
\end{equation}
such that
\begin{equation}\label{bsharp2}
b^{\sharp}\ |\psi_{i}\rangle=\sqrt{\rho_{i+1}}\
|\psi_{i+1}\rangle,
\end{equation}
so one can see that for pseudo-Hermitian system, biorthonormal
system, instate of $b^\dag$, $b^{\sharp}$ is the creation operator.
Using Eq. (\ref{bsharp2}) we get
\begin {equation}\label{coherent2}
|\theta\rangle_{n}=\sum_{i=0}^{n-1}\
\frac{\bar{q}^{^{\frac{i(i+1)}{2}}}}{{\rho_{i}!}}\ \theta {^{i}}\
({b^{\sharp}}){^{i}} |\psi_{0}\rangle,
\end {equation}
and considering to the following q-commutator relations
\begin{equation}\label{q-com}
\begin{array}{cc}
  {[\theta , b ^\sharp]}_{q}=0 & {[b, \theta }]_{q}=0 \\
  {[b ^\sharp , \bar{\theta}]}_{q}=0 & {[\bar{\theta} , b ]}_{q}=0 \\
\end{array}
\end{equation}
The coherent state derived above can be written in a compact form as:
\begin {equation}\label{coherent3}
|\theta\rangle_{n}=\sum_{i=0}^{n-1}\ \frac{({b^{\sharp}\
\theta}){^{i}}}{{\rho_{i}!}} |\psi_{0}\rangle=: e_{q}^ {(b^\sharp
\theta)}\ |\psi_{0}\rangle
\end {equation}
where the q-commutator and the generalized q-exponential function
are defined as:
\begin{equation}\label{comm exponen}
{[A,B]}_{q}:=AB-qBA, \qquad e_{q}^
{x}:=\sum_{n=0}\frac{x^n}{\rho_{n}!}.
\end{equation}
One must note that it is possible to construct another family of
GPHCS, $|\tilde{\theta}\rangle_{n}$, in terms of
$|\phi_{i}\rangle$. These coherent states are the eigen-basis of the
operator $\tilde{b}$ which annihilates the dual states
$|\phi_{i}\rangle$
\begin{equation}\label{tcoherent0}
\tilde{b}\ |\tilde{\theta}\rangle_{n} =\theta \
|\tilde{\theta}\rangle_{n},
\end{equation}
where the explicit form of the operator $\tilde{b}$ is defined as:
\begin{equation}\label{tilde{b}}
\tilde{b}=\eta b \eta^{-1}=\sum_{i=0}^{n-1}\sqrt{\rho_{i+1}}\
|\phi_{i}\rangle\langle\psi_{i+1}|,
\end{equation}
then the coherent state corresponding to the dual states
$|\phi_{i}\rangle$  can be obtained as follows
 \begin{equation}\label{tcoherent1}
|\tilde{\theta}\rangle_{n}=\sum_{i=0}^{n-1}\
\frac{\bar{q}{^{\frac{i(i+1)}{2}}}}{\sqrt{\rho_{i}!}}\ \theta
^{i}\ |\phi_{i}\rangle.
\end{equation}
Now in order to determine
$|\tilde{\theta}\rangle$ in terms of $|\phi_{0}\rangle$ let
have the following definition:
\begin{equation}\label{1}
 b^{\sharp '}:=\eta'^{-1}b^{\dag}\eta',\quad\quad where\quad \eta'^{-1}=\eta,
\end{equation}
according to the Eqs. (\ref{tilde{b}}) and (\ref{1}) we find   $ \tilde{b}^{\sharp '}= b^\dag$, and
then  $|\tilde{\theta}\rangle_{n}$ is:
\begin{equation}\label{tcoherent2}
|\tilde{\theta}\rangle_{n}=\sum_{i=0}^{n-1}\
\frac{\bar{q}^{^{\frac{i(i+1)}{2}}}}{{\rho_{i}!}}\ \theta
^{i}\ ({\tilde{b}^{\sharp '}})^{i} |\phi_{0}\rangle.
\end{equation}
The q-commutation relation between
$\theta$ and $\tilde{b}^{\sharp'}$ is:
\begin{equation}\label{q-com2}
{[\theta , \tilde{b}^{\sharp'}]}_{q}=0,
\end{equation}
then the Eq. (\ref{tcoherent2}) reduce to the following form
\begin{equation}\label{tcoherent3}
|\tilde{\theta}\rangle_{n}=\sum_{i=0}^{n-1}\
\frac{({\tilde{b}^{\sharp '}\ \theta})^{i}}{{\rho_{i}!}}
|\phi_{0}\rangle= e_{q}^ {(\tilde{b}^{\sharp '} \theta)}\
|\phi_{0}\rangle.
\end{equation}
which is the q-exponential form of $|\tilde{\theta}\rangle_{n}$. Now
we have all the ingredient to prove the completeness of the GPHCS
which is the task of next subsection.
\subsection{Resolution of Identity}
Let us  examine whether or not the GPHCS (
$|\theta\rangle_{n}$ and $|\tilde{\theta}\rangle_{n}$ ) satisfy
over-completeness property. By introducing  a weight function
\begin{equation}\label{weight}
w(\theta, \bar{\theta})=\sum_{i,j=0}^{n-1} \alpha_{ij}\theta^i
\bar{\theta}^j,
\end{equation}
and considering to the
Eqs. (\ref{qdeform1})-(\ref{thetapsiphi}), (\ref{q-com}) and
(\ref{q-com2}) it is clear that neither the integral
$|\theta\rangle\langle\theta|$ nor the integral of
$|\tilde{\theta}\rangle\langle\tilde{\theta}|$, against the
measure $d\bar{\theta}\ d\theta\  w(\theta, \bar{\theta})$
 unnormalized i.e.,:
\begin{equation}\label{resolution1}
\int d\bar{\theta}\ d\theta\  w(\theta, \bar{\theta})\
|\theta\rangle\langle\theta| \neq I, \qquad \int d\bar{\theta}\
d\theta\ w(\theta, \bar{\theta})\
|\tilde{\theta}\rangle\langle\tilde{\theta}|\neq I.
\end{equation}
 In order to overcome to this impasse and realized the resolution of identity  it is necessary to consider the Eq. (\ref{Hamiltonian})  i.e., the  biorthonormal nature of the system. Then it is reasonable to check the integrals $|{\theta}\rangle\langle\tilde{\theta}|$ and $|\tilde{\theta}\rangle\langle{\theta}|$  against the
measure $d\bar{\theta}\ d\theta\  w(\theta, \bar{\theta})$.
Choosing the proper weight function we can  realized the resolution of identity as:
\begin{equation}\label{resolution2}
\int d\bar{\theta}\ d\theta\  w(\theta, \bar{\theta})\
|\theta\rangle\langle\tilde{\theta}| = \int d\bar{\theta}\
d\theta\ w(\theta, \bar{\theta})\
|\tilde{\theta}\rangle\langle\theta|=I.
\end{equation}
To identify the weight function we replac the explicit form of $|\theta\rangle$ and
$\langle\tilde{\theta}|$ from Eqs. (\ref{coherent3}) and
(\ref{tcoherent3}) into Eq. (\ref{resolution2}) and we get
\begin{equation}\label{resolution3}
\int d\bar{\theta}\ d\theta\  w(\theta, \bar{\theta})\
|\theta\rangle\langle\tilde{\theta}| =\int d\bar{\theta}\ d\theta\
\sum_{k,l=0}^{n-1}c_{k,l}\theta^k
\bar{\theta}^l\sum_{i,j=0}^{n-1}\frac{\bar{q}^{\frac{i(i+1)}{2}}}{\sqrt{\rho_{i}!}}\
\theta^i|\psi_{i}\rangle\langle\phi_{j}|\ \bar{\theta}^j \
\frac{{q}^{\frac{j(j+1)}{2}}}{\sqrt{\rho_{j}!}}.
\end{equation}
Tacking account the quantization rules
(Eq. (\ref{gG}),(\ref{thetapsiphi})) and the integration rules of
generalized Grassmannian variables as in  Eq. (\ref{brezin}) it
becomes
\begin{equation}\label{resolution4}
=\sum_{k,l=0}^{n-1} \sum_{i,j=0}^{n-1} c_{k,l}\
\frac{\bar{q}^{\frac{i(i+1)}{2}}}{\sqrt{\rho_{i}!}}\
\
\frac{{q}^{\frac{j(j+1)}{2}}}{\sqrt{\rho_{j}!}}\ q^{j(i-1)}
\bar{q}^{j(j-1)} q^{il}\
\delta_{k+i,n-1}\delta_{l+j,n-1}|\psi_{i}\rangle\langle\phi_{j}|
\end {equation}
which in turn by the completeness of the biorthonormal basis
$\sum_{i=0}^{n-1}|\psi_{i}\rangle\langle\phi_{i}|=I $ of the
pseudo-Hermitian  system, then it leads to the following constrain
on the $c_{ij}$ coefficients :
\begin {equation}\label{cij}
c_{i,j}=\rho_{i}!\ q^{i(i+1)} \delta_{i,j}
\end{equation}
Thus the weight function must be equal
\begin{equation}\label{w}
w(\theta, \bar{\theta})=\sum_{i=0}^{n-1} q^{i(i+1)}\
\rho_{n-i-1}!\ \theta^i \bar{\theta}^i
\end{equation}
Finally  we find that the continuous set of  $|\theta\rangle$ and
$|\tilde{\theta}\rangle$ produce the system of biorthnormal
coherent state  which  provide a resolution of identity
(bi-over-completeness) for pseudo-Hermitian system.
\subsection{Time Evolution of GPHCS}
In this section we shall discuss the time evolution of the GPHCS.
The coherent states remain coherent for all the times provided that the time evolution of the initial state
$|\theta,0\rangle_{n} \equiv |\theta\rangle_{n}$, $|\theta, t\rangle_{n}$, managed by Hamiltonian, is also an eigen-state
 of lowering operator $b$
\begin{equation}\label{evolvedcoherentstate}
b\ |\theta,t\rangle_{n}=\theta(t)\ |\theta,t\rangle_{n},\quad where \quad|\theta,t\rangle_{n}=e^{-iHt}|\theta\rangle_{n}.
\end{equation}
Recalling $|\theta\rangle_{n}$ from Eq. (\ref{coherentform1}) one can write:
\begin{equation}\label{evolvedcoherentstate1}
\hspace{30mm}|\theta,t\rangle_{n}=\sum_{k=0}^{n-1}\
\frac{\bar{q}^{^{\frac{k(k+1)}{2}}}}{\sqrt{\rho_{k}!}}\ \theta
^{k}\ e^{-iE_{k}t}|\psi_{k}\rangle,
\end{equation}
to get the  proper solution suppose:
$$
E_{k}=-(n-k-2)E.
$$
Then one can express the
evolved coherent states as:
\begin{equation}
|\theta,t\rangle_{n}=e^{i(n-2)Et}|\theta(t)\rangle_{n}, \quad where \quad \theta(t)=e^{-iEt}\theta,
\end{equation}
thus, the evolved coherent states are actually remain eigen-states of annihilation operator,
which manifests the stability of the time evolution of CS
$|\theta\rangle$.
Similarly one can show that the $|\widetilde{\theta,t}\rangle$s are also stable and $|\theta,t\rangle$ and $|\widetilde{\theta,t}\rangle$
satisfy the resolution of Identity as
$$
\int d\bar{\theta}\ d\theta\  w(\theta, \bar{\theta})\
|\theta,t\rangle\langle\widetilde{\theta,t}| = \int d\bar{\theta}\
d\theta\ w(\theta, \bar{\theta})\
|\widetilde{\theta,t}\rangle\langle\theta,t|=I
$$
\subsection{$SU_{q}(2)$ Deformed Pseudo-Hermitian Coherent States}
In this section, we will derive the coherent states for the modified $SU(2)$ i.e., $SU_{q}(2)$,
following the technique developed in the previous section. For this purpose, consider a three level pseudo-Hermitian system,
we need first to start with these quantization relations:
\begin{equation}\label{}
\begin{array}{cc}
  \theta \ |\psi_{0}\rangle = \bar{q} \ |\psi_{0}\rangle\ \theta\quad , & \quad  \langle\psi_{0}|\ \bar{\theta} = q \ \bar{\theta}\ \langle\psi_{0}| \\
  \theta \ |\psi_{1}\rangle =  \ |\psi_{1}\rangle\ \theta \quad , & \quad  \langle\psi_{1}|\ \bar{\theta} =  \ \bar{\theta}\ \langle\psi_{1}| \\
  \theta\ |\psi_{2}\rangle = q \ |\psi_{2}\rangle \ \theta \quad , & \quad  \langle\psi_{2}|\ \bar{\theta} = \bar{q} \ \bar{\theta}\ \langle\psi_{2}| \\
  \bar{\theta} \ |\psi_{0}\rangle = q \ |\psi_{0}\rangle\ \bar{\theta}\quad , & \quad \theta\ \langle\psi_{0}|= q\langle\psi_{0}|\theta   \\
  \bar{\theta} \ |\psi_{1}\rangle =  \ |\psi_{1}\rangle\ \bar{\theta} \quad , & \quad \theta\ \langle\psi_{1}|= \langle\psi_{1}|\theta   \\
  \bar{\theta}\ |\psi_{2}\rangle = \bar{q} \ |\psi_{2}\rangle \ \bar{\theta} \quad , & \quad  \theta\ \langle\psi_{2}|= \bar{q}\langle\psi_{2}|\theta. \\
\end{array}
\end{equation}
The same relations hold between $|\phi_{i}\rangle$ and $\theta$ and
$\bar{\theta}$. Let us define the operators $b$, $b^{\sharp}$ and
$b_{z}$ as:
$$
b:= \sqrt{\rho_{1}}
|\psi_{0}\rangle\langle\phi_{1}|+\sqrt{\rho_{2}}
|\psi_{1}\rangle\langle\phi_{2}|,
$$
$$
b^{\sharp}:= \eta^{-1}b^{\dag}\eta= \sqrt{\rho_{1}}
|\psi_{1}\rangle\langle\phi_{0}|+\sqrt{\rho_{2}}
|\psi_{2}\rangle\langle\phi_{1}|,
$$
$$
b_{z}:=[b,b^{\sharp}]_{q}:=bb^{\sharp}-q b^{\sharp}b.
$$
Using the explicit form of the operators $b$, $b^{\sharp}$ and $b_{z}$ we will try to find
the conditions which make it possible that these operators be the generators of $su_{q}(2)$ Lie algebra.
To do so consider the commutation relation between $b$ and  $b_{z}$
\begin{equation}\label{bzb}
[b_{z},b]_{q}=(\rho_{1}-q \rho_{2}+q^2
\rho_{1})\sqrt{\rho_{1}}|\psi_{0}\rangle\langle\phi_{1}|+(\rho_{2}-q
\rho_{1}+q^2
\rho_{2})\sqrt{\rho_{2}}|\psi_{1}\rangle\langle\phi_{2}|,
\end{equation}
to get the proper solution  we have:
$$
(\rho_{1}-q \rho_{2}+q^2 \rho_{1})=(\rho_{2}-q \rho_{1}+q^2\rho_{2}),
$$
or equivalently
\begin{equation}\label{eqvalentely}
(1+q+q^2)(\rho_{1}-\rho_{2})=0 \Rightarrow q=e^{\frac{2\pi i}{3}},
\end{equation}
with the above restriction the Eq. (\ref{bzb}) reduces to the following forms:
\begin{equation}\label{reduceform}
{[b_{z},b]}_{q}=(\rho_{1}-q \rho_{2}+q^2
\rho_{1}) b=(\rho_{2}-q \rho_{1}+q^2
\rho_{2}) b.
\end{equation}
The same condition, Eq. (\ref{eqvalentely}), is required for the commutator of $b^{\sharp}$ and $b_{z}$ to satisfy the $su_{q}(2)$ algebra then we have:
\begin{equation}\label{suq2}
\left\{\begin{array}{c}
  \hspace{-7cm}[b,b^{\sharp}]_{q}=b_{z} \\
 \hspace{-.5cm} {[b_{z},b]}_{q}=(\rho_{1}-q \rho_{2}+q^2
\rho_{1}) b=(\rho_{2}-q \rho_{1}+q^2
\rho_{2}) b \\
  {[b^{\sharp},b_{z}]}_{q}=(\rho_{1}-q \rho_{2}+q^2
\rho_{1})b^{\sharp} =(\rho_{2}-q \rho_{1}+q^2 \rho_{2})b^{\sharp}
\end{array}\right.
\end{equation}
Considering the Eq. (\ref{coherent}), the coherent states in question
are:
$$
b\ |\theta\rangle =\theta \ |\theta\rangle
$$
and $|\theta\rangle$ can be found as:

\begin{equation}\label{squ2coherent}
\begin{array}{c}
  |\theta\rangle= |\psi_{0}\rangle+\frac{\bar{q}}{\sqrt{\rho_{1}}}\
\theta \ |\psi_{1}\rangle+ \frac{1}{\sqrt{\rho_{1}\rho_{2}}}\
\theta^2 \ |\psi_{2}\rangle \\
\;\;\;\;\; = \big(1+\frac{\bar{q}}{\sqrt{\rho_{1}}}\
\theta \ b^\sharp + \frac{1}{\sqrt{\rho_{1}\rho_{2}}}\
\theta^2 \ {b^\sharp}^2 \big)|\psi_{0}\rangle \\
\end{array}
\end{equation}
where in the second equality in Eq. (\ref{squ2coherent}) we have used the operator $b^\sharp$ to introduce  $|\theta\rangle$ in terms of $|\psi_{0}\rangle$
and finally based on the definition of the q-exponential function, Eq. (\ref{comm exponen}), we have:
$$
|\theta\rangle= e_{q}^ {(b^\sharp \theta)}\ |\psi_{0}\rangle.
$$
To get the dual space coherent states we recall the $\tilde{b}$
$$
\tilde{b}=\eta b \eta^{-1}=\sqrt{\rho_{1}}
|\phi_{0}\rangle\langle\psi_{1}|+\sqrt{\rho_{2}}
|\phi_{1}\rangle\langle\psi_{2}|,
$$
the dual space coherent states are the eigen states of $\tilde{b}$  :
$$
\tilde{b}\ |\tilde{\theta}\rangle =\theta \ |\tilde{\theta}\rangle,
$$
then the dual GPHCS states are:
\begin{equation}\label{tildeteta}
|\tilde{\theta}\rangle=
|\phi_{0}\rangle+\frac{\bar{q}}{\sqrt{\rho_{1}}}\ \theta \
|\phi_{1}\rangle+ \frac{1}{\sqrt{\rho_{1}\rho_{2}}}\ \theta^2 \
|\phi_{2}\rangle=\eta\ |\theta\rangle,
\end{equation}

and using the the definition of q-exponential function, Eq. (\ref{comm exponen}), we get:
\begin{equation}
|\tilde{\theta}\rangle= e_{q}^ {(\tilde{b}^{\sharp'} \theta)}\
|\psi_{0}\rangle.
\end{equation}
\subsection{Resolution of Identity}
This section is treats the over completeness of the generalized
Grassmannian coherent state for three level system. Considering the
Eq. (\ref{weight}) the weight function of three level system is:
$$
w(\theta, \bar{\theta})=\sum_{i,j=0}^2 \alpha_{ij}\theta^i
\bar{\theta}^j.
$$
Taking into account the pseudo-Hermiticity of the system, both of the following integrals
do not satisfy the resolution of identity i.e.,
$$
\int d\bar{\theta}\ d\theta\  w(\theta, \bar{\theta})\
|\theta\rangle\langle\theta| \neq I, \qquad \int d\bar{\theta}\
d\theta\ w(\theta, \bar{\theta})\
|\tilde{\theta}\rangle\langle\tilde{\theta}|\neq I
$$
but like that of the GPHCS, Eq. (\ref{resolution2}), we can get the over completeness for the integrals:
$$
\int d\bar{\theta}\ d\theta\  w(\theta, \bar{\theta})\
|\theta\rangle\langle\tilde{\theta}| = \int d\bar{\theta}\
d\theta\  w(\theta, \bar{\theta})\
|\tilde{\theta}\rangle\langle\theta|=I
$$
with the following proper weight function:
$$
w(\theta, \bar{\theta})=\rho_{1}\rho_{2}+\frac{\rho_{1}}{q}\
\theta \bar{\theta}+ \theta^2 \bar{\theta}^2
$$
\subsection{Stability}
Turning to the results of section 3.4 we will study the stability of generalized Grassmannian coherent states of $SU_{q}(2)$.
As mentioned in section 3.4, time evolution of coherent states governed by Hamiltonian and the evolved coherent states will be coherent states if they remain as the eigen states of annihilation operator i.e., Eq. (\ref{evolvedcoherentstate}).
Now considering Eq. (\ref{evolvedcoherentstate}) and the explicit form of $SU_{q}(2)$ coherent states i.e, Eq. (\ref{squ2coherent})
the evolved coherent state is :
\begin{equation}\label{evolveSUq2}
|\theta,t\rangle=e^{-i E_{0}
t}|\psi_{0}\rangle+\frac{\bar{q}}{\sqrt{\rho_{1}}}\ e^{-i E_{1}
 t}\theta\ |\psi_{1}\rangle+ \frac{1}{\sqrt{\rho_{1}\rho_{2}}}\
e^{-i E_{2} t}\theta^2 \ |\psi_{2}\rangle
\end{equation}
Taking into account that $E_{0}=-E$ , $E_{1}=0$ and $E_{2}=E$, we
put $\theta(t)=e^{-iEt}\theta$ and rewrite the above equation in
the form
$$
|\theta,t\rangle=e^{i E
t}\big(|\psi_{0}\rangle+\frac{\bar{q}}{\sqrt{\rho_{1}}}\
\theta(t)\ |\psi_{1}\rangle+ \frac{1}{\sqrt{\rho_{1}\rho_{2}}}\
\theta^{2}(t) \ |\psi_{2}\rangle\big)=e^{i E t}\ |\theta(t)\rangle
$$
which manifests the stability of the time evolution of CS
$|\theta\rangle$.
\subsection{Squeezed states}
Based on the definition of standard squeezing operator we define generalized Grassmannian pseudo-Hermitian
squeezing operator as follows:
\begin{equation}\label{squeez}
S(\theta)=\exp[\frac{1}{2}(\theta b^{\sharp^2}-\ \bar{\theta}b^2)],
\end{equation}
it is easy to check that for three level system, $SU_{q}(2)$
algebra, the operators $b^{3}$ and $b^{\sharp^3}$ are zero then the
squeezing operator is reduced to:
\begin{equation}\label{squoper}
S(\theta)=I+\frac{1}{2}(\theta b^{\sharp^2}-\ \bar{\theta}
b^2)-\frac{\bar{q}}{4}\ \theta\bar{\theta}\
(b^{\sharp^2}b^2+qb^2b^{\sharp^2}).
\end{equation}
The generalized Grassmannian pseudo-Hermitian squeezed states by definition are obtained from the application of the
$S(\theta)$ on the ground state, $|\psi_{0}\rangle$, i.e.:
\begin{equation}
|\xi\rangle=S(\theta)|\psi_{0}\rangle.
\end{equation}
Using Eq. (\ref{squoper}) we have
$$
|\xi\rangle=|\psi_{0}\rangle+\frac{\sqrt{\rho_{1}\rho_{2}}}{2}\theta|\psi_{2}\rangle
-\frac{\rho_{1}\rho_{2}}{4}\theta\bar{\theta}|\psi_{0}\rangle,
$$
\begin{equation}\label{squ state}
\;\;\;\;=(1-\frac{\rho_{1}\rho_{2}}{4}\theta\bar{\theta})|\psi_{0}\rangle+
\frac{\sqrt{\rho_{1}\rho_{2}}}{2}\theta|\psi_{2}\rangle.
\end{equation}
similar to the case of GPHCS, there is another family of generalized Grassmannian pseudo-Hermitian squeezed states
which can be obtained from the action of $\eta$ on $|\xi\rangle$, Eq. (\ref{tildeteta}),
then
\begin{equation}\label{tildexi}
|\tilde{\xi}\rangle=\eta|\xi\rangle=(1-\frac{\rho_{1}\rho_{2}}{4}\theta\bar{\theta})|\phi_{0}\rangle+
\frac{\sqrt{\rho_{1}\rho_{2}}}{2}\theta|\phi_{2}\rangle.
\end{equation}
Finally we define the operator $\tilde{S}(\theta)$ in terms of
$S(\theta)$ as follow
\begin{equation}\label{p-squeez}
\tilde{S}(\theta)=\eta S(\theta)
\eta^{-1},
\end{equation}
using the Eq. (\ref{squoper}), make it possible to rewrite $\tilde{S}(\theta)$ as:
$$\tilde{S}(\theta)=I+\frac{\sqrt{\rho_{1}\rho_{2}}}{2}\big(\
\theta|\phi_{2}\rangle\langle\psi_{0}|-\bar{\theta}
|\phi_{0}\rangle\langle\psi_{2}|
\big)-\frac{{\bar{q}\rho_{1}\rho_{2}}}{4}\theta\bar{\theta}\big
(|\phi_{2}\rangle\langle\psi_{0}|+q
|\phi_{0}\rangle\langle\psi_{2}|\big),
$$
and considering the fact that for $SU_{q}(2)$ algebra the operators $\tilde{b}^{\sharp'^3}$ and $\tilde{b^3}$
are zero then we can get the exponential form  of $\tilde{S}(\theta)$ as:
$$
\tilde{S}(\theta)=\exp[{\frac{1}{2}(\theta \tilde{b}^{\sharp'^2}-\
\bar{\theta} \tilde{b}^2)}].
$$
\section{Conclusion}
Generalized  Grassmannian coherent states associated to the pseudo-Hermitian
lowering operator, which annihilate the eigen-basis of $n$
level pseudo-Hermitian  Hamiltonian $H$ are constructed. Taking to
account the bi-orthonormal nature of the pseudo-Hermitian system, it
is possible to prepare two families of the coherent states.
Meanwhile resolution of identity is discussed and it is explored
that, the resulted coherent states  satisfy the
bi-over-completeness condition instead of over-completeness one
i.e., their bi-over-completeness inherited from the
bi-orthonormality of the pseudo-Hermitian eigen-basis. Furthermore
as a special case the generalized Grassmannian coherent states of
q-deformed $SU(2)$, $SU_{q}(2)$, for $3$ level system were studied in
detail and finally the squeezed states of this three level system
were introduced.
Finally we note that the construction of GPHCS outlined here may also be
extended to compound systems of the fermionic and Grassmannian density matrix in quantum information theory.

\end{document}